\documentclass{emulateapj}

\received{}
\revised{}
\accepted{}
\shorttitle{Numerical Model} \shortauthors{Peer \& Waxman}

\newcommand{\beq}{\begin{equation}}
\newcommand{\eeq}{\end{equation}}
\newcommand{\ba}{\begin{array}}
\newcommand{\ea}{\end{array}}

\newcommand{\cm}{\mbox{ cm}}
\newcommand{\eV}{\mbox{ eV }}
\newcommand{\keV}{\mbox{ keV }}
\newcommand{\MeV}{\mbox{ MeV }}
\newcommand{\GeV}{\mbox{ GeV }}
\newcommand{\TeV}{\mbox{ TeV }}

\def \etal{{\it et al.~}}

\begin{document}
\title {Time dependent numerical model for the emission of radiation
from relativistic plasma}  
\author{Asaf Pe'er\altaffilmark{1}\altaffilmark{2} and Eli Waxman\altaffilmark{1}}
\altaffiltext{1}{Department of Condensed Matter Physics, Weizmann
Institute, Rehovot 76100, Israel}
\altaffiltext{2}{asaf@wicc.weizmann.ac.il}

% --------------------------------------------------------------------------
% Version Notes:
% First draft, July 10th, 2003. 
% V2.0: Start Jan. 11th, 2004
% V3.1: Aug. 17th, 2004 (After Eli corrections)
% --------------------------------------------------------------------------

\begin{abstract}

We describe a numerical model constructed for the study of the
emission of radiation from relativistic plasma under conditions
characteristic, e.g., to gamma-ray bursts (GRB's) and active galactic
nuclei (AGN's). The model solves self consistently the kinetic
equations for $e^\pm$ and photons, describing cyclo-synchrotron
emission, direct Compton and inverse Compton scattering, pair
production and annihilation, including the evolution of high energy
electromagnetic cascades. 
The code allows calculations over a wide range of particle energies,
spanning more than 15 orders of magnitude in energy and time scales.
Our unique algorithm, which enables to follow the particle
distributions over a wide energy range, allows to accurately derive
spectra at high energies, $>100 \TeV$. 
We present  the kinetic equations that are being solved, detailed
description of the equations describing the various physical processes,
the solution method, and several examples of numerical results. 
Excellent agreement with analytical results of the synchrotron-SSC
model is found for parameter space regions in which this approximation
is valid, and several examples are presented of calculations for
parameter space regions where analytic results are not available.

\end{abstract}

\keywords{galaxies: active --- gamma rays: bursts ---
 gamma rays:theory  --- methods: numerical --- 
plasmas --- radiation mechanism: Non thermal
}

\section{Introduction}
\label{sec:intro}
\indent
In the standard fireball scenario of gamma-ray bursts (GRB's), the
observable effects are due to the dissipation of kinetic energy in a
highly relativistic fireball  (see, e.g.,
\citet{fireballs1,fireballs2,W03} for reviews). 
Synchrotron emission and inverse-Compton emission
by accelerated electrons are the main radiative processes.  Electrons
accelerated in the internal shock waves within the expanding fireball
produce the prompt $\gamma$-ray emission, while electrons accelerated
in the external shock wave driven by the fireball into the surrounding
medium produce the afterglow emission, from the X to the radio bands.
\citep{PR93, MR97, Viet97, Sari98, GW99}.

While being in general agreement with observations
\citep{Band93, Preece98, Frontera00,fireballs2}
% \citep[prompt emission:][]{Band93, Preece98, Frontera00},
% \citep[afterglow:][]{Akerlof, Costa97, Van97}, and see also \citep{fireballs2}
both theoretical arguments and observational evidence suggest that the
 optically thin synchrotron - synchrotron self Compton (SSC) emission
 model is not complete in explaining neither the prompt nor the
 afterglow emission.   
Additional physical processes can significantly modify the SSC spectrum.
First, over a wide range of model parameters, a large number of
 $e^\pm$ pairs are produced in internal collisions, due to
 annihilation of high energy photons.
Second, relativistic pairs cool rapidly to mildly-relativistic
energy, where their energy distribution is determined by a balance
between emission and absorption of radiation. The emergent spectrum,
which is affected by scattering off the pair population, depends
strongly on the pair energy distribution, and in particular on the
"effective temperature" which characterizes the low-end of the energy
distribution.
Third, proton and electron acceleration to high energies initiates rapid
 electro-magnetic cascades. It is necessary to follow the evolution of
 a high energy, non-linear cascade in order to accurately derive the
 spectrum.
And last, the plasma is not in steady state, and the particle distributions
 are continuously evolving. 

On the observational side, we note that hard spectra, $\nu F_\nu
\propto \nu^\alpha$ with $\alpha > 4/3$ at low energies, $\lesssim
300 \keV$, were observed at early times in some GRB's \citep{Preece98,
  Frontera00, Ghirlanda03}.  
These hard spectra is inconsistent with the optically thin
synchrotron-SSC model predictions. A similar conclusion was obtained
by a comparison of the high energy and low energy spectral indices during
the prompt emission phase of 150 GRB's \citep{Preece02}. 
An additional  high energy ($> 1 \MeV$) spectral component inconsistent with the
synchrotron model prediction was reported by \citet{Gonzalez}.
Finally, a  recent analysis by
\citet{BB04} showed difficulty in explaining the high energy component
of GRB's early emission by the SSC model.

The above mentioned difficulties, raise the need for a model that
  can better describe emission under conditions characterizing
  GRB's. However, a numeric calculation of GRB spectra that takes into consideration
creation and annihilation of pairs is complicated. 
The evolution of electromagnetic cascades initiated by the
annihilation of high energy photons occurs on a very short time scale. 
On the other extreme, evolution of the low-energy,  mildly
relativistic pairs, which is governed by synchrotron  self-absorption,
direct and inverse Compton emission takes much longer time. 
The large difference in characteristic time scales poses a challenge
to numeric calculations.
Another challenge to numerical modeling is due to the fact that at
mildly relativistic energies the usual synchrotron emission and IC
scattering approximations are not valid, and a precise
cyclo-synchrotron emission, direct Compton and inverse Compton
scattering calculations are required.

Two approaches have been employed so far in treating this problem. The
first is the Monte-Carlo approach, where individual particles are
followed as they undergo interactions inside the plasma.
This scheme typically suffers from relatively poor photon statistics at 
high energies, and does not lend itself to time-dependent calculations.
Work done so far using this approach \citep{Pilla98} was limited to
parameter space region where the creation of pairs has only a minor
effect on the resulting spectrum. 
The second approach involves solving the relevant kinetic equations. 
Following the time evolution of the system using this method is
straightforward, and photon statistics at high energies is not an
issue. However, the above mentioned complications limited the accuracy
of the numerical models constructed so far \citep{PM98} above $\sim 1
\GeV$.

Note that this method was extensively used in the past in the study of
active galactic nuclei (AGN) plasma \citep{ZL85, FBGPC,
LZ87, Coppi92}.  Using the numerical models new results were obtained,
such as the effective pair temperature and the complex pattern of the
spectral indices in the X-ray ($2-10 \keV$) range \citep{LZ87}, that
were not obtained by previous analytic calculations.
Non of these models, however, considered the evolution of high energy
electro-magnetic cascades, expected to be relevant for both GRB's 
and AGN's. In addition, the treatment of photon emission in
the presence of magnetic field was not complete, since particles are
expected to accumulate at low energies ($\gamma \sim 1$), where the
synchrotron emission approximation used does not hold, and exact
treatment of cyclo-synchrotron emission is required.

Pair cascade evolution was first studied by \citet{BR71}. Small angle
cascade showers in anisotropic radiation field were treated by
\citet{BL82}. \citet{Guilbert83} and \citet{Svensson87} have
generalized the treatment of cascade evolution, showing that it may
have a significant effect on the high energy spectrum. It is therefore
necessary to incorporate the cascade calculation in order to
accurately derive the high energy emission spectrum. 
 
We have constructed a numerical model that overcomes the numerical
challenges. Applying this model to GRB plasmas, we have obtained
several new results. For example, we have shown \citep{Pe'er03b} that
emission peaks at $\sim 1\MeV$ for $\tau_\pm \sim 10 - 10^2$, where
$\tau_\pm$ is the optical depth to scattering by pairs, and that peak
energy at $\gg \MeV$ cannot be obtained for GRB luminosity $ L \sim
10^{52} \rm{\, ergs \, s^{-1}}$. We showed that for large compactness,
$l' > 100$, the spectral slope below $ 1 \MeV$ is steep,
$\varepsilon^2 n_{ph}(\varepsilon) \propto \epsilon^\alpha$ with $0.5 <
\alpha < 1$ and shows a sharp cutoff at $10 \MeV$.
We also showed  \citep{Pe'er03} that observations of the early
afterglow emission at $1 \GeV - 1 \TeV$ is informative about two of
the most poorly determined parameters of the fireball model: the ambient
matter density, and the fraction of thermal energy carried by the
magnetic field, $\epsilon_B$.

We present in this paper our numerical model.
In \S\ref{sec:procesess} we describe the basic model assumptions. We
then present the kinetic equations that are being solved, and detailed
description of the numerical treatment of various physical processes. 
Our numerical integration approach is described in \S\ref{sec:numerical}.
We present the general approach of treating this complicated problem,
and the various integration techniques used.
In \S\ref{sec:results} we give examples of numerical results, relevant
to the prompt emission phase of GRB's, and compare them to approximate
analytic results. We summarize in \S\ref{sec:summary} the main
features of our method, and discuss its usefulness for the ongoing
research of GRB's and AGN's.

\section{Model assumptions and physical processes}
\label{sec:procesess}
\indent
We consider a uniform plasma, composed of protons, electrons,
positrons and photons, and permeated by a time independent magnetic
field. The particle and photon distributions are assumed homogeneous
and isotropic.  
Considering the physical phenomenon of, e.g., GRB as a motivation,
these assumptions are equivalent to the assumption that the
calculations are carried out in the comoving frame (see
\S\ref{sec:results} below). 
We assume the existence of a dissipation process, (e.g., collisionless
shock waves) which produces energetic particles at constant rates,
$Q(\gamma)$ and $S(\gamma)$ electrons and protons respectively per
unit time per unit volume per unit Lorentz factor, $\gamma$. 
Since the details of the acceleration process are not yet known, we do
not specify here the functions $Q(\gamma)$ and $S(\gamma)$. These
functions are specified when treating a particular problem (see
\S\ref{sec:results}).
Motivated by the GRB fireball model scenario, in which the internal shock
waves cross the colliding shells at relativistic speeds, we assume
that the dissipation process occurs on a characteristic time scale
which is equal to the light crossing time, $t_{dyn} \sim R/c$, where
$R$ is a characteristic length scale of the plasma.
   
The population of electrons, positrons and photons is affected by
synchrotron emission, synchrotron self absorption, Compton scattering,
pair production and pair annihilation, that occur simultaneously
during the dynamical time, $t_{dyn}$.  In this version of the code,
protons are assumed to interact via photo-meson interactions only,
producing pions that decay into energetic photons and positrons.
Coulomb scattering is not considered, because, as we show in
(\S\ref{coulomb}), it is insignificant in calculating the spectra 
under conditions that are of interest to us. As noted by
\citet{Coppi90}, $e-e$ bremsstrahlung is also insignificant under the
same conditions, and is therefore not included in the calculations.   

We assume no photon escape during the dynamical time, $t_{dyn}$, and
instantaneous photon release at the end of the dynamical time. This
approximation is justified since the dynamical time is equal to the
light crossing time. 
If $\tau_T$, the optical depth to Thompson scattering by electrons or by the
created pairs at the end of the dynamical time is larger than 1, the
'instantaneous release' approximation is not valid.
Since the plasma is assumed to be heated to relativistic energy
density, we assume in this case that the dissipation phase is followed
by a relativistic expansion phase, during which the optical depth
decreases. The evolution of particle and photon distributions is
followed during the expansion phase until the optical depth for Thompson
scattering, $\tau_T<1$. A detailed description of this calculation is
given in \citet{Pe'er03b}. 

Let $n_{e^-}(\gamma,t)$, $n_{e^+}(\gamma,t)$ and $n_p(\gamma,t)$ be the
number density per unit Lorentz factor, $\gamma$, per unit volume of
electrons, positrons and protons, and $n_{ph}(\varepsilon,t)$ be the
number density per unit energy per unit volume of photons.
The time derivatives of the electron, positron, proton and photon number
densities are given by 
%\beq
\begin{eqnarray}
\frac{\partial n_{e^-}(\gamma,t)}{\partial t} & =&  Q(\gamma) + 
\frac{\partial}{\partial \gamma} \Bigg[ n_{e^-}(\gamma,t) \left( P_S(\gamma)
+ P_C(\gamma,t) \right)  \nonumber \\
& &  + H(\gamma,t) \beta \gamma^2
\frac{\partial}{\partial \gamma} \left( \frac{n_{e^-}(\gamma,t)}{\beta
\gamma^2} \right) \Bigg]  \nonumber \\
& &+ \frac{\partial n_{eP}(\gamma,t)}{\partial t} -  
\frac{\partial n_{eA}(\gamma,t)}{\partial t}, 
\label{eq:master,el}
\end{eqnarray}
%\eeq
\begin{eqnarray}
\frac{\partial n_{e^+}(\gamma,t)}{\partial t} & = &
\frac{\partial}{\partial \gamma}\Bigg[ n_{e^+}(\gamma,t) \left( P_S(\gamma)
+ P_C(\gamma,t) \right) \nonumber \\ & & + H(\gamma,t) \beta \gamma^2
\frac{\partial}{\partial \gamma} \left( \frac{n_{e^+}(\gamma,t)}{\beta
\gamma^2} \right) \Bigg]  \nonumber \\
& & + \frac{\partial n_{eP}(\gamma,t)}{\partial t} -  
\frac{\partial n_{eA}(\gamma,t)}{\partial t} + Q_\pi(\gamma,t), 
\label{eq:master,pos}
\end{eqnarray}
\beq
\frac{\partial n_p(\gamma,t)}{\partial t} = S(\gamma) + 
\frac{\partial}{\partial \gamma}\left[n_p(\gamma,t) P_\pi(\gamma,t) \right],
\label{eq:master,pr}
\eeq
\beq 
\ba{rcl}
\frac{ \partial n_{ph}(\varepsilon,t)}{\partial t} & = & R_S(\varepsilon,t)
+ R_C(\varepsilon,t) - R_P(\varepsilon,t) +  R_A(\varepsilon,t)
\nonumber \\ & &
+ R_\pi(\varepsilon,t) - c n_{ph}(\varepsilon,t) \alpha(\varepsilon,t). 
\label{eq:master,ph}
\ea
\eeq 
Here, the terms in the parenthesis on the right hand side of equations
\ref{eq:master,el},\ref{eq:master,pos}, give the change of population
due to synchrotron emission and Compton scattering, $P_S(\gamma)$ and
$P_C(\gamma,t)$ are the synchrotron and Compton emitted power. The third term in
the parenthesis represents energy gain by synchrotron self absorption,
with $H(\gamma,t)$ defined below, see equation \ref{eq:H}. The last two terms
in equation \ref{eq:master,el}, $\partial n_{eP}(\gamma,t)/\partial t$
and $\partial n_{eA}(\gamma,t)/\partial t$ are the rates of pair
creation and pair annihilation per unit volume. 
The term $Q_\pi(\gamma,t)$ in equation \ref{eq:master,pos} represents
positron creation by the decay of $\pi^+$. The pions are produced by
photo-meson interactions of low energy photons with energetic protons.  
In the proton equation, $P_\pi(\gamma,t)$ is the rate of proton energy
transfer to pions.
In the photon equation, $R_S(\varepsilon,t)$, $R_C(\varepsilon,t)$,
$R_P(\varepsilon,t)$ and $R_A(\varepsilon,t)$ are the rate of production
and annihilation of photons due to synchrotron emission, Compton
scattering, pair production and pair annihilation, and
$R_\pi(\varepsilon,t)$ is the production rate of photons due to the decay of
energetic pions. 
The last term represents photons reabsorption, where
$\alpha(\varepsilon,t)$ is the self absorption coefficient.

In the rest of this section, we present detailed description of the
terms in equations \ref{eq:master,el}--\ref{eq:master,ph}.

\subsection{Synchrotron and synchrotron self absorption emission terms}
\label{sec:ssa}
\indent

The term $H(\gamma,t)$ in equation \ref{eq:master,el} describes the
heating of the electrons and their diffusion in energy due to
synchrotron self absorption. It is given by 
\beq
H(\gamma,t) = \int d \omega \frac{I_\omega(t)}{4 \pi m_e \omega^2} P(\omega,\gamma)
\label{eq:H}
\eeq    
\citep[see][]{McCray69,Ginzburg69,GGS88}.  
The specific intensity, $I_\omega(t)$, is calculated using $I_\omega(t) =
n_{ph}(\varepsilon,t) \varepsilon c \hbar / 4 \pi$, where $\varepsilon
=\hbar \omega$. $P(\omega,\gamma)$ is the total power emitted by an
electron having Lorentz factor $\gamma$ per unit frequency $\omega$,
and is given in \S\ref{sec:sync.}.

The time derivative of photon distribution due to synchrotron emission
is given by 
\beq
R_S(\varepsilon,t) = \frac{1}{\hbar}\int d \gamma
P(\omega,\gamma) n_{e^\pm}(\gamma,t),
\eeq
where $n_{e^\pm}(\gamma,t) \equiv n_{e^-}(\gamma,t) + n_{e^+}(\gamma,t)$.

In a homogeneous plasma, the self absorption coefficient is given by 
\beq
\alpha(\varepsilon,t) = - \frac{\pi^2}{8  m_e \omega^2}
\int d\gamma P(\omega, \gamma) \beta \gamma^2 \frac{\partial}{\partial \gamma}
\left[ \frac{n_{e^\pm}(\gamma,t)}{\beta \gamma^2}\right]
\eeq
\citep{Ginzburg69,Rybicki}.

\subsubsection{Cyclo-Synchrotron emission spectrum}
\label{sec:sync.}
\indent
The power (energy/time/sr/frequency) emitted by a single
electron moving with velocity $\beta \equiv v/c$ in a frequency range
$\omega .. \omega + d \omega$, 
at an angle $\theta$ with respect to the magnetic field, 
is given by
\beq
\ba{lcl}
\eta_\omega({\mathbf \beta},\theta) d\omega & = & 
\frac{q^2 \omega^2}{2 \pi c}  \Bigg[
\sum_{m=1}^{\infty}\left( \frac{\cos \theta - \beta_\parallel}{\sin \theta}
\right)^2 J_m^2(x)  \nonumber \\ & & + \beta_\perp^2 {J'}_m^2(x) 
\Bigg]
\delta(y) d\omega
\ea
\eeq
\citep[see][]{Bekefi66, Ginzburg69, Mahadevan96}.
Here
\beq
x = \frac{\omega}{\omega_0} \beta_\perp \sin \theta,
\eeq

\beq
\omega_0 = \frac{\omega_b}{\gamma} ; \quad 
\omega_b \equiv \frac{q B}{m_e c}, 
\eeq

\beq
y = m \omega_0 - \omega(1- \beta_\parallel \cos \theta).
\eeq
$J_m(x)$ is the Bessel function of order m, $J'_m(x)$ its derivative,
$\beta_\parallel = \beta \cos \theta_p$ and 
$\beta_\perp = \beta \sin \theta_p$  are the velocity components
parallel and perpendicular to the magnetic field, and $\theta_p$ is 
the angle between the electron velocity direction and the 
magnetic field.
The presence of a $\delta$~-function implies that the emission occurs
at discrete frequencies.

The total power emitted by a single electron having Lorentz factor
$\gamma \equiv (1-\beta^2)^{-1/2}$ per unit frequency $\omega$ is given
by integrating $\eta_\omega({\mathbf \beta},\theta)$ over the solid
angle $d\Omega \equiv \sin \theta d\theta d\phi$.
For an isotropic distribution of electrons, the mean radiated power is
given by  
%\begin{eqnarray}
\beq
\ba{rcl}
P(\omega,\gamma) \equiv \frac{dE}{dt d\omega} & = &\frac{2}{4\pi}
\int_0^{2\pi} d \phi_p \int_0^1 d(\cos \theta_p) \nonumber \\
& & \times
\int_0^{2\pi} d\phi \int_{-1}^1 d(\cos \theta) 
\eta_\omega({\mathbf \beta}, \theta) \nonumber \\ 
& = & 2\pi  \int_0^1 d(\cos \theta_p)
\int_{-1}^1 d(\cos \theta) 
\eta_\omega({\mathbf \beta}, \theta),
\label{eq:P_omega_general}
\ea
\eeq
%\end{eqnarray}
where the factor of $1/4 \pi$ comes from angular normalization of the
isotropic distribution, and the factor of 2 is due to integration on
half of the range of $\cos \theta_p$.

In the synchrotron limit $\gamma \gg 1$,
the Bessel functions can be approximated by modified 
Bessel functions, resulting in the well known result
\beq
P(\omega,\gamma) = \frac{\sqrt{3} q^3 B \sin \theta_p}{2 \pi m_e c^2} F(X),
\label{eq:P_omega_sync}
\eeq 
where
\beq
X = \frac{\omega}{\omega_c}, \quad
\omega_c = \frac{3}{2} \gamma^2 \frac{q B} {m_e c} \sin \theta_p.
\eeq
$F(X)$ is given by
\beq
F(X) \equiv X \int_X^\infty K_{5/3}(\xi) d\xi,
\eeq  
where $K_{5/3}(\xi)$ is modified Bessel function.
The function $F(X)$ was tabulated in, e.g., \citet{Ginzburg65}.

The power emitted by a single electron is given by 
integrating eqs. \ref{eq:P_omega_general}, \ref{eq:P_omega_sync}
over all frequencies,
\beq
P_S(\gamma) \equiv \int P(\omega,\gamma) d \omega = 
\frac{ 2 q^4 B^2 \gamma^2 \beta^2 \sin^2 \theta_p}{3 m_e^2 c^3}
\label{eq:P_gamma}
\eeq
\citep[see, e.g.,][]{Rybicki}.

The calculation method of the cyclo-synchrotron emission spectrum
$P(\omega,\gamma)$, is determined by the electron energy:
(i) For low energy electrons, having  $\gamma<3.2$
($\beta < 0.95$), integration of equation 
\ref{eq:P_omega_general} is carried out explicitly, at all frequencies
up to $\omega/\omega_b = 200$. 
Above this frequency, no emission is assumed.
(ii) For electrons with $3.2 < \gamma <10$, equation
\ref{eq:P_omega_general} is solved up to  $\omega/\omega_b = 100$. 
Above this frequency, the approximate synchrotron spectrum 
(eq. \ref{eq:P_omega_sync}) is calculated up to $\omega \leq 10 \omega_c$.
(iii) At high electron energies, $\gamma > 10$ the synchrotron spectrum 
(eq. \ref{eq:P_omega_sync}) is calculated in the range
$0.001 \omega_c < \omega < 10 \omega_c$.

\subsection{Compton scattering}
\label{sec:Compton}
\indent

The total power emitted  by Compton scattering by a single electron
having Lorentz factor $\gamma$ into a unit volume, is given by  
\beq
P_C(\gamma,t) = \int d \alpha_1 \int d \alpha
\frac{d^2N(\gamma,\alpha_1)}{dt d\alpha} n_{ph} (\alpha_1,t) (\alpha -
\alpha_1),
\label{eq:P_c}
\eeq
where $d^2N(\gamma,\alpha_1)/dtd\alpha$ is the rate of scattering by 
a single electron having Lorentz factor $\gamma$ passing through space
filled with a unit density (1 photon per unit volume), isotropically distributed,
mono-energetic photons with energy $\alpha_1 m_ec^2$.
Note that the Compton power can be negative (i.e., the electron gains
energy), depending on the initial photon number density distribution,
$n_{ph}(\alpha_1,t)$. 

The time evolution of the photon number density due to Compton
scattering is given by
\beq
\ba{rcl}
R_C(\alpha,t) & = & \int d \gamma n_{e^\pm}(\gamma,t) \nonumber \\
& & \times
\int d \alpha_1 n_{ph}(\alpha_1,t) \left[
\frac{d^2N(\gamma,\alpha_1)}{dt d\alpha} - 
\frac{d^2N(\gamma,\alpha)}{dt d\alpha_1} \right].
\ea
\eeq

%========================
%
%========================
\subsubsection{Compton scattering spectrum}
\label{sec:Compton_spectrum}

The rate of scattering by a single electron having Lorentz factor
$\gamma$ passing through space filled with a unit density,
isotropically distributed, mono-energetic photons with
energy $\alpha_1 m_ec^2$ was first derived by \citet{Jones68},
\beq
\frac{d^2N(\gamma,\alpha_1)}{dt d\alpha} = 
\frac{ \pi r_0^2 c \alpha}{2 \gamma^4 \beta \alpha_1^2}
\left[F(\zeta_+) - F(\zeta_-) \right].
\label{eq:compton_rate_general}
\eeq
Here, $\alpha$ is the energy of the outgoing photon in units of
$m_e c^2$, $r_0$ is the classical electron radius, $\beta =
(1-1/\gamma^2)^{1/2}$, $\zeta_\pm$
are the upper and lower integration limits (see below)
and $F(\zeta)$ is given by the sum of 12 functions obtained 
by solving equation (21) of \citet{Jones68}
\footnote{Note that in eq. 21 of Jones, there is a misprint by a
  factor of $a$ in the one before last term.}\footnote{
Note, though, that \citet{Coppi90} claim about an error in eq. (20) of 
 \citet{Jones68} is incorrect. In fact,  eq. A1.1 of \citet{Coppi90}
is identical to eq. (20) of  \citet{Jones68}.}.

Solving equation (21) of \citet{Jones68}, $F(\zeta)$ is given by
\beq
\begin{array}{l}
F(\zeta) = \sum_{i=1}^{12} f_i, \\ 
f_1 = \left(\frac{\gamma}{\alpha}\right)^2  
\left(\frac{\gamma}{\alpha_1}\right) \sqrt{E_1}, \\
f_2 = -\left(\frac{\gamma}{\alpha}\right)
\frac{2}{\sqrt{a}} \log\left( \frac{\sqrt{a} + \sqrt{E_1}}{\sqrt{b \zeta}}
\right), \\
f_3 = - \frac{\sqrt{E_1}}{a \zeta} - 
\frac{\alpha_1}{\gamma}\frac{2}{a^{3/2}}
 \log\left( \frac{\sqrt{a} + \sqrt{E_1}}{\sqrt{b \zeta}}\right), \\
f_4 = - \left(\frac{\gamma}{\alpha}\right)^2 
\left( \alpha_1/\gamma+1 \right) 
\left( \alpha/\alpha_1 +1 \right)
\frac{1}{\sqrt{E_1}}, \\
f_5 = \left(\frac{\gamma}{\alpha}\right)^2 \frac{\gamma}{2 \alpha_1}
\left( \sqrt{E_1} + \frac{a}{\sqrt{E_1}} \right), \\
f_6 = \left(\frac{\gamma}{\alpha}\right)
\left(\alpha_1/\gamma +1\right)^2 \frac{2}{a \sqrt{E_1}}
-2 \left(\frac{\gamma}{\alpha}\right) \frac{(\alpha_1/\gamma +1)^2}{a^{3/2}}
 \log\left( \frac{\sqrt{a} + \sqrt{E_1}}{\sqrt{b \zeta}}\right), \\
%\end{array}
%\eeq
%\beq
%\begin{array}{l}
f_7 =  -4 \left(\frac{\gamma}{\alpha}\right) 
\frac{\gamma} {\sqrt{|c|}}
\left\{
\begin{array}{ll}
\sinh^{-1} \left(\sqrt{c \zeta / d}\right) & c>0; \\
\sin^{-1} \left(\sqrt{-c \zeta / d}\right) & c<0,  \\
\end{array}
\right. \\

f_8 = \left(\frac{\gamma}{\alpha}\right)^2 \gamma \sqrt{E_2} / c
-  \left(\frac{\gamma}{\alpha}\right)^2 d 
\frac{\gamma}{|c|^{3/2}}
\left\{
\begin{array}{ll}
\sinh^{-1} \left(\sqrt{c \zeta / d}\right) & c>0; \\
(-1) \sin^{-1} \left(\sqrt{-c \zeta / d}\right) & c<0,  \\
\end{array}
\right. \\
f_9 = -\frac{ 2 \gamma} {d} \frac {\sqrt{E_2}}{\zeta}, \\
f_{10} = \frac{ 4 \alpha c \zeta}{d^2 \sqrt{E_2}} + 
\frac{2 \alpha}{d \sqrt{E_2}}, \\
f_{11} = \alpha \gamma^2 \left( \frac{\alpha_1}{\gamma} -
 \frac{\alpha}{\gamma}+1+\frac{\alpha_1}{\alpha} \right) 
\frac{2 \zeta}{d \sqrt{E_2}},\\
f_{12} = \alpha_1 \gamma^2 \frac{2 \zeta}{c \sqrt{E_2}} -
 \alpha_1 \gamma^2 \frac{2}{|c|^{3/2}}
\left\{
\begin{array}{ll}
\sinh^{-1} \left(\sqrt{c \zeta / d}\right) & c>0; \\
(-1) \sin^{-1} \left(\sqrt{-c \zeta / d}\right) & c<0.  \\
\end{array}
\right. \\
\end{array}
\label{eq:sum_f}
\eeq
Here, 
\beq
\begin{array}{l}
a = 1/\gamma^2 \left[ \left( \alpha_1 + \gamma \right)^2-1 \right], \\
b = 2 \alpha_1/\gamma , \\
c = (\gamma - \alpha)^2 -1 , \\
d = 2 \alpha/\gamma , \\
E_1 = a - b \zeta , \\
E_2 = c \zeta^2 + d \zeta .\\
\end{array}
\eeq

The integration limits depend on the energy of the outgoing
photon, $\alpha$, best presented as a function of the parameter
$\rho \equiv \alpha/\alpha_1$.
The minimum value of $\rho$ is\footnote{Note that there is a misprint
in the result that appears in \citet{Jones68}.}
\beq
\rho_{\min} = \frac{1-\beta}{1+\beta+2\alpha_1/\gamma}, 
\eeq
while the upper value of $\rho$ is limited by the kinematics,
\beq
\rho_{\max,1} =  1 + (\gamma-1)/\alpha_1,
\eeq
and by the requirement that $\zeta \leq 1+\beta$,
\beq
\rho_{\max,2} = \frac{1+\alpha_1/\gamma+\sqrt{(1+\alpha_1/\gamma)^2
-1 +\beta^2-2\alpha_1/\gamma(1+\beta)}}
{1-\beta+2\alpha_1/\gamma},
\eeq
resulting in $\rho_{\max} = \min(\rho_{\max,1}, \rho_{\max,2})$.

For a given $\rho$, $\rho_{\min} \leq \rho \leq \rho_{\max}$,
the integration boundaries are 
\beq
\ba{rcl}
\zeta_-(\rho) & = & \max \Big(\rho \Big\{(1+\alpha_1/\gamma -
\rho\alpha_1/\gamma) \nonumber \\ & & 
- \left[(1+\alpha_1/\gamma -  \rho \alpha_1/\gamma)^2-1/\gamma^2\right]^{1/2}
\Big\}, 1-\beta \Big),
\ea
\eeq
and
\beq
\ba{rcl}
\zeta_+(\rho) & = & \min \Big(\rho \Big\{(1+\alpha_1/\gamma -
\rho\alpha_1/\gamma) \nonumber \\ & &
+ \left[(1+\alpha_1/\gamma -  \rho \alpha_1/\gamma)^2-1/\gamma^2\right]^{1/2}
\Big\}, 1+\beta \Big).
\ea
\eeq

For an energetic electron, $\gamma \gg 1$ and $\gamma \gg \alpha_1$, 
equation \ref{eq:compton_rate_general} can be simplified
and the scattering rate is given by \citep{Jones68,BG70} 
\beq
\ba{rcl}
\frac{d^2N(\gamma,\alpha_1)}{dt d\alpha} & \approx & 
\frac{ 2 \pi r_0^2 c}{\alpha_1 \gamma^2}
\Bigg[2 q \log q + (1+2q)(1-q) \nonumber \\ & &  +\frac{1}{2}
\frac{(4 \alpha_1 \gamma q)^2}{(1+4\alpha_1\gamma q)}(1-q) \Bigg],
\label{eq:compton_rate_limit}
\ea
\eeq
where $q \equiv \alpha/4 \alpha_1 \gamma^2(1-\alpha/\gamma)$
is limited to $1/4\gamma^2 < q\leq 1$.

Calculation of the spectrum resulting from Compton scattering is
determined by the electron Lorentz factor $\gamma$ and the incoming
photon energy, $\alpha_1$. 
(i) For $\gamma > 10^4$ and $\alpha_1<10^{-5}$, the
approximate spectrum (eq. \ref{eq:compton_rate_limit}) is used.
(ii) For all other values of $\gamma, \alpha_1$, the exact spectrum
(eq. \ref{eq:compton_rate_general}) is calculated. 
The results are tabulated in a 3-d matrix (initial electron energy
$\times$ initial photon energy $\times$ final photon energy), and are
used in calculating the time derivatives of electron and photon
number densities.

%========================
%
%========================
\subsection{Pair production}
\label{sec:pair_prod}
\indent
  
The production rate of particles having Lorentz factor in the range
$\gamma .. \gamma + d\gamma$
by an isotropic photon field with photon density $n_{ph}(\alpha,t)$,
was calculated by \citet{BS97},
\beq
\ba{rcl}
{\partial n_{eP}(\gamma,t) \over \partial t} & = & \frac{3}{4}\sigma_T c
\int_0^\infty d\alpha_1  
\frac{n_{ph}(\alpha_1,t)}{\alpha_1^2} \nonumber \\ & & \times
\int_{\max\left\{\frac{1}{\alpha_1},\gamma+1-\alpha_1\right\}}^\infty
d\alpha_2 \frac{n_{ph}(\alpha_2,t)}{\alpha_2^2} \nonumber \\ & & \times
\left.
\left\{ \frac{\sqrt{E^2-4\alpha_{cm}^2}}{4} + H_+ + H_- \right\}
\right|_{\alpha_{cm}^L}^{\alpha_{cm}^U}.
\label{eq:pair_prod_full}
\ea
\eeq
$\alpha_{1,2}$ are the
scattering photons energies in units of $m_ec^2$, $E = \alpha_1 +
\alpha_2$, and $\alpha_{cm}$ is the photons energy in the center
of momentum frame, given by $ 2 \alpha_{cm}^2 = \alpha_1 \cdot
\alpha_2$. The functions $H_\pm$ are calculated using
\beq
c_\pm \equiv (\alpha_{1,2} - \gamma)^2 - 1,
\eeq
\beq
d_\pm \equiv \alpha_{1,2}^2 + \alpha_1 \alpha_2 
\pm \gamma (\alpha_2 - \alpha_1).
\eeq
For $c_\pm \neq 0$, $H_\pm$ are given by
\begin{eqnarray}
H_\pm & = &  - \frac{\alpha_{cm}}
{8\sqrt{\alpha_1 \alpha_2 + c_\pm \alpha_{cm}^2}}
\left(\frac{d_\pm}{\alpha_1 \alpha_2} + \frac{2}{c_\pm} \right) 
\nonumber \\
& & + \frac{1}{4}\left(2- \frac{\alpha_1 \alpha_2 -1}{c_\pm}\right)
I_\pm
\nonumber \\ & & 
+ \frac{\sqrt{\alpha_1 \alpha_2 + c_\pm \alpha_{cm}^2}}{4}
\left(\frac{\alpha_{cm}}{c_\pm} + 
\frac{1}{\alpha_{cm} \alpha_1 \alpha_2} \right),
\end{eqnarray}
where
\beq
I_\pm = \left\{
\begin{array}{ll}
\frac{1}{\sqrt{c_\pm}} \ln \left(\alpha_{cm}\sqrt{c_\pm} + 
\sqrt{\alpha_1 \alpha_2 + c_\pm \alpha_{cm}^2} \right) & c_\pm >0 , \\
\frac{1}{\sqrt{-c_\pm}} \arcsin 
\left( \alpha_{cm}\sqrt{-\frac{c_\pm}{\alpha_1 \alpha_2}} \right) &
c_\pm < 0 . \\
\end{array} \right.
\eeq
For $c_\pm = 0$,
\beq
\ba{rcl}
H_\pm & = & \left(\frac{\alpha_{cm}^3}{12} - 
\frac{\alpha_{cm}d_\pm}{8} \right) \frac{1}{(\alpha_1 \alpha_2)^{3/2}}
\nonumber \\ & & 
+ \left(\frac{\alpha_{cm}^3}{6} + \frac{\alpha_{cm}}{2}
+ \frac{1}{4 \alpha_{cm}} \right) \frac{1}{\sqrt{\alpha_1 \alpha_2}}.
\ea
\eeq

The upper and lower integration limits $\alpha_{cm}^U$, $\alpha_{cm}^d$ are given by
\beq
\alpha_{cm}^U = \min \left\{ \sqrt{\alpha_1 \alpha_2}, \alpha_{cm}^a
\right\}, \nonumber \\
\alpha_{cm}^L = \max \left\{ 1, \alpha_{cm}^b \right\},
\eeq
where
\beq
\left(\alpha_{cm}^{a,b}\right)^2 = \frac{1}{2} \left(
\gamma [E-\gamma] + 1 \pm \sqrt{(\gamma[E-\gamma]+1)^2 -E^2} \right).
\eeq

The total loss rate of photons in the energy range
$\alpha_1.. \alpha_1 + d\alpha_1 $  
by pair production is given by
\beq
\ba{rcl}
R_P(\alpha_1,t) & = & - n_{ph}(\alpha_1,t) \frac {c}{2}
 \int d (\cos \theta) ( 1 - \cos \theta) \nonumber \\ & & \times
\int_{\frac{2}{\alpha_1 (1-\cos \theta)}}^\infty
 d\alpha_2 n_{ph}(\alpha_2,t)
\sigma (\alpha_1, \alpha_2, \theta),
\label{eq:prod_rate} 
\ea
\eeq
where
\beq 
\ba{rcl}
\sigma(\alpha_1, \alpha_2, \theta)& = & \frac{3}{16} \sigma_T
(1-\beta'^2) \nonumber \\ 
& & \times \left[ 2\beta'(\beta'^2-2)+
(3-\beta'^4)\ln\left(\frac{1+\beta'}{1-\beta'}\right)\right],
\ea
\eeq
and
\beq
\beta' = \left[ 1- \frac{2}{\alpha_1 \alpha_2 (1-\cos \theta)} 
\right]^{1/2}
\eeq
\citep{Gould-Schreder67, Lang99}.
The resulting particle spectra are symmetric for electrons and positrons.

Calculation of the photon loss rate is carried out using equation
\ref{eq:prod_rate}. The spectra of the emergent pairs is calculated in
accordance to the photons energies: 
(i)  For $1.001 \leq \alpha_1 \cdot \alpha_2 \leq 10^4$, equation
\ref{eq:pair_prod_full} is solved, and the exact spectrum is obtained.
(ii) For $\alpha_1 \cdot \alpha_2 < 1.001$, monoenergetic
spectrum of the created particles assumed, with energy $(\alpha_1+\alpha_2)/2$.
(iii) If $\alpha_1 \cdot \alpha_2 > 10^4$, one of the created
particles energy is taken to be $\alpha_{\max} \equiv \max
(\alpha_1,\alpha_2)$, and for the second particle the energy is
approximated as $\alpha_{\min} + 1/(2 \alpha_{\min})$, where 
$\alpha_{\min} \equiv \min (\alpha_1,\alpha_2)$.

\subsection{Pair Annihilation}
\label{sec:pair_ann}
\indent
The total loss rate of electrons having Lorentz factor $\gamma_1
.. \gamma_1 + d\gamma_1$ due to pair production (in the plasma frame)
is given by 
\beq
\ba{rcl}
{{\partial n}_{e^-A}(\gamma_1,t) \over \partial t} & = & -
n_{e^-}(\gamma_1,t) \frac{c}{2\gamma_1} 
\int d (\cos \theta)\nonumber \\ & & \times \int d \gamma_2
n_{e^+}(\gamma_2,t)  
\beta'_2 \frac{dn'}{dn} \sigma_{ann}(\gamma'_2),
\label{eq:ann_rate}
\ea
\eeq
where 
$\gamma'_2 = \gamma_1 \gamma_2 ( 1 + \beta_1 \beta_2 \cos \theta)$ is 
the positron Lorentz factor in the electrons rest frame, $\beta'_2$
is its velocity in this frame and 
$dn'/dn = \gamma_1(1+\beta_1\beta_2\cos\theta)$.
The cross section for a positron having Lorentz factor $\gamma$
to annihilate with an electron at rest, $\sigma_{ann}(\gamma)$, is
given by 
\beq
\ba{rcl}
\sigma_{ann}(\gamma) & = & \frac{3}{8} \frac{\sigma_T}{\gamma+1}
\nonumber \\ & & \times \left[
\frac{\gamma^2 + 4\gamma + 1}{\gamma^2 -1} 
\ln(\gamma + \sqrt{\gamma^2-1})-\frac{\gamma+2}{\sqrt{\gamma^2-1}}
\right]
\ea 
\eeq
\citep{Svensson82, Lang99}.
The loss rate of positrons is calculated in a similar way.

The annihilation rate is calculated by solving equation \ref{eq:ann_rate}.
Since (i) It was shown in \citet{Svensson82} that for a large
region of $\gamma_1$, $\gamma_2$ the photons spectrum is narrowly
peaked around $\varepsilon_{1,2} = \gamma_{1,2} m_e c^2$, and
(ii) We found numerically that calculation of the exact particle
spectrum resulting after pair production, compared to the approximate
particles spectrum $\gamma_{1,2} = \varepsilon_{1,2} / m_e c^2$, did not have a
significant effect on the resulting photon spectra, we decided not to 
include calculation of the pair annihilated photon spectra in this
version of the code. The emergent photons energies assumed to be equal to
the reacting particles energies, $\varepsilon_{1,2} = \gamma_{1,2} m_e
c^2$,
thus
\beq
R_A(\epsilon = \gamma m_e c^2,t) =
- {{d n}_{eA}(\gamma,t) \over dt}.
\eeq

%===================

\subsection{Photon and positron production by $\pi$ decay}
Photo-meson interactions between energetic protons  and low energy
photons result in production of $\pi$'s. The fractional energy loss
rate of a proton with Lorentz factor $\gamma_p$ due to pion production is 
\beq 
\ba{rcl}
t_\pi^{-1}(\gamma_p,t) & \equiv & - {1 \over \gamma_p} {d \gamma_p
  \over dt} \nonumber \\
& = & \frac{1}{2\gamma_p^2}c \int_{\varepsilon_0}^\infty 
d\varepsilon \sigma_\pi (\varepsilon) \xi (\varepsilon) \varepsilon  
\int_{\varepsilon/2 \gamma_p}^\infty dx x^{-2} n_{ph}(x,t), 
\label{eq:t_pi1} 
\ea
\eeq 
\citep{WB97}
where $\sigma_\pi(\varepsilon)$ is the cross section for pion production 
for a photon with energy $\varepsilon$ in the proton rest frame, 
$\xi(\varepsilon)$ is the average fraction of energy lost to the pion, 
and $\varepsilon_0 = 0.15 \GeV$ is the threshold energy. 
For a flat photon spectrum ($\varepsilon^2 n_{ph}(\varepsilon) \propto
\varepsilon^\alpha$ with $\alpha \simeq 0$), the contribution to the
first integral of equation \ref{eq:t_pi1} from photons at the
$\Delta$-resonance is comparable to that of photons of higher energy,
thus 
\beq 
t_\pi^{-1}(\gamma_p,t) = \frac{c}{\gamma_p^2} 
\Delta \varepsilon \sigma_{peak} \xi_{peak} \varepsilon_{peak}
\int_{\varepsilon_{peak}/2 \gamma_p}^\infty dx x^{-2} n_{ph}(x,t), 
\label{eq:t_pi2} 
\eeq 
where $\sigma_{peak} \simeq 5 \times 10^{-28} \rm{\, cm^2}$ and
$\xi_{peak} \simeq 0.2$ at the resonance 
$\varepsilon = \varepsilon_{peak} = 0.3 \GeV$, and $\Delta \varepsilon \simeq 0.2$
is the peak width.

The rate of proton energy transfer to pions is given by
\beq
P_\pi(\gamma,t) = t_\pi^{-1}(\gamma_p,t)\times \gamma_p m_p c^2.
\eeq 
The energy loss rate of protons is calculated by numerical integration
of the integral in equation \ref{eq:t_pi2}. This calculation is carried out only in
those cases where the $\Delta$-resonance approximation is valid, and
can easily be extended to any photon spectrum by explicit integration
of the integrals in equation \ref{eq:t_pi1}.
Roughly half of this energy is converted into high energy photons
through the  $\pi^0$ decay. Each of the created photons carry 10\% of
the initial proton energy, thus the photon production rate is given by
\beq
R_\pi(\varepsilon = \gamma_p m_p c^2 / 10,t) = 5
t_\pi^{-1}(\gamma_p,t) n_p(\gamma_p,t),
\label{eq:R_pi}
\eeq
where $n_p(\gamma_p)$ is the number density of protons at energy
$\gamma_p m_p c^2$.
Half of the energy lost by protons is converted into $\pi^+$, that
decays into positron and neutrinos, $\pi^+ \rightarrow \mu^+ + \nu_\mu
\rightarrow e^+ + \nu_e + \bar{\nu}_\mu + \nu_\mu$. The $\pi^+$'s
energy is roughly evenly distributed between the decay products, thus
the positron carries 5\% of the initial proton energy, and the
positron production rate is given by
\beq
Q_\pi(\gamma = \gamma_p (m_p/m_e) / 20,t) = 2.5 t_\pi^{-1}(\gamma_p,t)
n_p(\gamma_p,t).
\label{eq:Q_pi}
\eeq

Equations \ref{eq:R_pi}, \ref{eq:Q_pi} provide only a crude
approximation to the spectrum of high energy photons and positrons
produced by pion decay. However, photons and positrons that are
created by pion decay are typically very energetic, and participate in
the high energy electro-magnetic cascade. Since these particles and
photons' energy is spread among the cascade products, and the final
cascade spectrum has only weak dependence on the initial spectrum, it
is appropriate to use the approximate expressions in equations
\ref{eq:R_pi} and \ref{eq:Q_pi}. 

%===================

%===================

%===================

\section{Numerical approach}
\label{sec:numerical}
\indent
Several integration methods are used in solving the kinetic equations.
Simple, first order difference scheme was found adequate, except when
dealing with synchrotron self absorption and with the evolution of the
rapid high energy electro-magnetic cascades. Synchrotron self
absorption calculations are carried using Cranck-Nickolson second
order integration scheme \citep[see][]{NR}.

The particle and photon distributions are discretized, spanning the
energy range relevant to the problem. Note that in the problems
involved, this energy range can extend over 20 decades (see
\S\ref{sec:results} below). 
Spectra of cyclo-synchrotron emission (eqs. \ref{eq:P_omega_general},
\ref{eq:P_omega_sync}), Compton scattering
(eqs. \ref{eq:compton_rate_general},  \ref{eq:compton_rate_limit}) and
pair production (eq. \ref{eq:pair_prod_full}) are pre-calculated and
stored in tables.  

Following simultaneously the evolution of the rapid high energy
electro-magnetic cascade and the much slower evolution of low energy
processes, is difficult.
The 'stationary' approximation used in previous works in treating the
evolution of high energy particles \citep[see, e.g.,][] {FBGPC, LZ87,
Coppi92} can not be used, due to the non-linear nature of the cascade:
As an energetic particle loses its energy, many secondaries are
created, which, in turn, serve as primaries for further development of
the cascade. As the cascade evolution is powered by inverse Compton
scattering, pair production and annihilation, the injection rate of
energetic photons and pairs depends on the entire particle and photon
spectra.  

Therefore, in treating this problem, a fixed time step is chosen,
typically $10^{-4.5}$ times the dynamical time. 
Numerical integration is carried out with this fixed time step.
At each time step, the cascade evolution is followed directly. 
%The calculation includes (i) The energy loss
%time of electrons and positrons at various energies (via
%cyclo-synchrotron emission and inverse Compton scattering taking into
%account the fact that low energy electrons can gain energy via direct
%Compton scattering); (ii) The annihilation time of electrons and
%positrons; (iii) The annihilation and energy loss time of photons. 
Direct numerical integration is carried out only for the
electrons, positrons and photons for which the energy loss
time or annihilation time are longer than the fixed time step. 
Electrons, positrons and photons for which the energy loss time or
 annihilation time are shorter than the fixed time step, are assumed
 to lose all their energy in a single time step, producing secondaries.
The secondaries' spectra are determined by the spectra of the various
 physical processes, as presented in \S\ref{sec:procesess}, and by the
 relative rates of these processes.
We discriminate between high energy secondaries, which are secondaries for
which the energy loss time or annihilation time are shorter than the
fixed time step, and low energy secondaries which lose their energy
or annihilate on a time scale longer than the fixed time step. 
The calculation is repeated for the high energy secondaries, which are 
 treated as a source of lower energy particles, until all the cascade
energy is transferred to low energy particles. 
Since in each step of the cascade calculation, part of the energy is
 transferred into low energy particles which do not participate in the
 cascade, convergence is guaranteed. 
In order to check for convergence of this method, we repeat the complete
calculation with a shorter time step. 

%The approximation used is justified for an integration time step which
% is much shorter than the dynamical time scale.
 
The time derivative of particle distributions due to synchrotron 
emission and Compton scattering is calculated by solving the
continuity equation, ${\partial n}(\gamma,t)/\partial t = \partial
j(\gamma,t) / \partial E$, where $j(\gamma,t) \equiv n(\gamma,t) P(\gamma)$, 
and $P(\gamma)$ is the emitted power (see eq. \ref{eq:master,el}).
In solving this equation, flux limiter is used to ensure convergence
for large time steps, and Neumann boundary conditions for the flux
$j(\gamma,t)$ at the boundary points are used. The rate of change of
particle distributions due to pair production and annihilation are
calculated using eqs. \ref{eq:pair_prod_full}, \ref{eq:ann_rate}.
Conservation of particle number and energy is forced using Lagrange
multiplier method. This method was found to allow faster
convergence (larger time steps).

At the low end of the particle spectrum, electrons and positrons gain
energy via direct Compton scattering, on a timescale shorter than the
fixed time step. In parallel to gaining energy, these particles lose
energy via synchrotron emission on a much longer time scale, thus
providing another challenge to numerical integration.
Defining 'very low energy particles' as particles that gain energy on
a time scale shorter than the fixed time step, we treat this problem
in the following way. At each time step, calculation of the number
density of these particle is repeated iteratively, until the particle
distribution converges, and the total emissivity equals the absorption.
At each of the iteration steps, the calculated emissivity and
absorption are stored, and used in the calculation of the photon
emission from these particles.   
Convergence of this method as well is checked by repeating the
calculation with smaller time step.  

%Toward the end of the dynamical time, shorter integration time steps
%are used, in order to smooth the obtained results.
% The calculation is repeated with a shorter time step, to guarantee
%convergence. 

\section{Examples of numerical results}
\label{sec:results}

We give below several examples of the results of numerical
calculations of GRB prompt emission spectra. Detailed description of
numerical results of prompt emission spectra and early afterglow
emission spectra are found in \citet{Pe'er03b,Pe'er03}.
Our calculations are done in the framework of the fireball model
\citep[see, e.g.,][]{fireballs1,fireballs2,W03},
where the emission results from electron acceleration to
ultra-relativistic energies by internal shocks within an expanding
wind.  

\subsection{Basic assumptions, plasma conditions and particle
  acceleration}
\label{sec:assumptions}
We calculate the emergent spectra following a single collision between
two plasma shells.  
Denoting by $\Gamma$ the characteristic wind Lorentz factor, and
assuming variation $\Delta \Gamma /\Gamma \sim 1$ on a time scale
$\Delta t$, two shells collide at radius $ r_i = 2\Gamma^2 c \Delta t$.
For $\Delta \Gamma /\Gamma \sim 1$, two mildly relativistic
($\Gamma_s-1\sim1$ in the wind frame) shocks are formed, one
propagating forward into the slower shell ahead, and one 
propagating backward (in the wind frame) into the faster shell behind.  
The comoving shell width, measured in the shell rest frame, 
is $\Delta R=\Gamma c \Delta t$, and the comoving dynamical time, 
the characteristic time for shock crossing and shell expansion measured 
in the shell rest frame, is $t_{dyn}=\Gamma\Delta t$.
The shock waves, which propagate at relativistic velocity $v_s \sim c$
in the plasma rest frame, dissipate the plasma kinetic energy and
accelerate particles to high energies. Since the shock velocity is
time independent during $t_{dyn}$, the shock-heated comoving plasma
volume is assumed to increase linearly with time, i.e., constant
particle number density is assumed.  

Under these assumptions, the shocked plasma conditions are determined
by six free parameters.  
Three are related to the underlying source: the total luminosity $L =
10^{52}\,L_{52} \rm{\ erg\,s^{-1}}$, the Lorentz factor of the shocked
plasma, $\Gamma = 10^{2.5}\,\Gamma_{2.5}$, and the variability time
$\Delta t = 10^{-4}\, \Delta t_{-4} \rm{\,s}$. 
Three additional parameters are related to the collisionless-shock
microphysics: The fraction of post shock thermal energy carried by
electrons $\epsilon_e = 10^{-0.5} \epsilon_{e,0.5}$ and by magnetic
field, $\epsilon_B = 10^{-0.5} \epsilon_{B,0.5}$, and the power law
index of the accelerated electrons' Lorentz factor distribution, $d\log n_e / d
\log \gamma = -p$ assumed to extend over the range $\gamma_{\min} \leq
\gamma \leq \gamma_{\max}$.   

The comoving proton number density is
\beq
n_p \approx \frac{L}{4 \pi r_i^2 c \Gamma^2 m_p c^2} = 
6.7 \times 10^{14} \ L_{52} \  \Gamma_{2.5}^{-6} \ 
 \Delta t_{-4}^{-2} \  \cm^{-3}.
\label{eq:n_p}
\eeq
The internal energy density is $u_{int} = L/(4 \pi r_i^2 c \Gamma^2)$,
resulting in a magnetic field, 
\beq
B = \sqrt{\frac{\epsilon_B L}{2 \Gamma^6 c^2 \Delta t ^2}} = 
2.9\times 10^6 \ L_{52}^{1/2} \  \epsilon_{B,-0.5}^{1/2} \ 
 \Gamma_{2.5}^{-3} \  \Delta t_{-4}^{-1} \  \rm{G}.
\label{eq:B}
\eeq

\subsubsection{Particle acceleration}
\label{acceleration}

Since the details of the acceleration mechanism are not yet known,
we adopt the common assumption of a power law energy distribution of
the accelerated electrons Lorentz factor, $\gamma$. The power law index
$p$ of the accelerated particles is a free parameter of the model.
The maximum Lorentz factor of the accelerated electrons,
$\gamma_{\max}$ is obtained by equating the acceleration time,
$t_{acc} = \gamma m_e c^2 /c q B$, and the synchrotron cooling time, 
$t_{syn} = 9 m_e^3 c^5 / 4 q^4 B^2 \gamma$, to obtain $\gamma_{\max} 
= \left({6 \pi q}/{\sigma_T B}\right)^{1/2}$.
The accelerated particles assume a power law energy distribution above a 
minimum Lorentz factor $\gamma_{\min}$, which is obtained by simultaneously solving 
\beq
\begin{array}{c}
n_e = \int_{\gamma_{\min}}^{\gamma_{\max}} \frac{dn_e}{d\varepsilon}
d\varepsilon, \nonumber \\ 
u_{e} =  \int_{\gamma_{\min}}^{\gamma_{\max}} \varepsilon
\frac{dn_e}{d\varepsilon} d\varepsilon, 
\end{array}
\label{eq:gamma_min}
\eeq
where $n_e$ and $u_e \equiv \epsilon_e u_{int}$ are the number and
energy densities of the electrons.
The injected particle distribution below $\gamma_{\min}$ is assumed
thermal with temperature $\theta \equiv kT/m_e c^2 = 3 \gamma_{\min}$,
and exponential cutoff is assumed above $\gamma_{\max}$. 
In the results shown below, no proton acceleration is assumed. 

Our calculations are carried in the plasma (comoving) frame.
The particle distributions are discretized, spanning total of 10
decades of energy, ($\gamma \beta_{\min} = 10^{-3}$ to $\gamma
\beta_{\max} = 10^{7}$).
The photon bins span 14 decades of energy, from $\alpha_{\min} \equiv 
\varepsilon_{\min}/m_ec^2 = 10^{-8}$ 
to $\alpha_{\max} \equiv \varepsilon_{\max}/m_ec^2 = 10^{6}$.
No a-priori photon field is assumed.

\subsection{Low compactness}
\label{sec:low_l}
\begin{figure}
\plotone{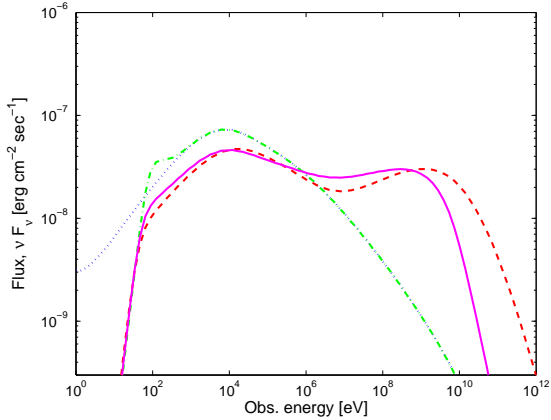}
\caption{Time averaged GRB prompt emission spectra obtained after 2
shells collision, characterized by low compactness parameter.
Results are shown for $L=10^{52}$~erg,
$\epsilon_e=\epsilon_B=10^{-0.5}$, $p=3$, $\Delta t = 10^{-2} {\rm \
s}$ , $\Gamma = 300$. The comoving compactness parameter is $l' = 2.5$.
Dotted curve: cyclo- synchrotron emission only. Dash dotted curve:
synchrotron emission and self absorption only. Dashed: synchrotron
emission, synchrotron self absorption and Compton scattering. Solid:
All processes included, including pair production and annihilation,
but excluding proton acceleration.
Luminosity distance $d_L=2 \times 10^{28}$ and $z=1$ were assumed.
}
\label{fig:results1}
\end{figure}

We examined the dependence of the emergent spectrum on the uncertain
values of the free parameters of the model.
We found that the spectral shape strongly depends on the
dimensionless compactness parameter $l$, defined by  $l
\equiv L \sigma_T / R m_e c^3$, where  $L$ is the luminosity, and $R$
is a characteristic length of the object.  
For low value of the comoving compactness, $l'\lesssim 10$, the
optical depth to pair production and to scattering by pairs is smaller
than $1$ \citep{Pe'er03b}, thus synchrotron-SSC emission model
provides fairly good approximation of the resulting spectrum.
Therefore, before applying our model to examine realistic
  scenario, (i.e., comparison with observations), we first compare our
  numerical results to the analytical model predictions, in
  parameter space region where the later are valid. 
%We therefore use this {\bf analytical} model predictions to test the
%numerical results.     
Figure \ref{fig:results1} presents numerical results in this case,
where a power low index $p=3$ was used to allow the synchrotron
and Compton peaks to be distinctively apparent.

The synchrotron peak presented in Figure~\ref{fig:results1} at
$\varepsilon_{peak}^{ob.} \simeq 10 \keV$ is 
in excellent agreement with the analytical results of the
  optically thin synchrotron model prediction,  
\beq 
\ba{rcl}
\varepsilon_{peak}^{ob.} & = & \hbar \frac{3}{2} \frac{q B}{m_e c}
 \gamma_{\min}^2 \Gamma \nonumber \\ & = & 1.4 \times 10^4 \quad   
L_{52}^{1/2} \  \epsilon_{e,-0.5}^2 \  \epsilon_{B,-0.5}^{1/2} \
\Gamma_{2.5}^{-2} \ \Delta t_{-2}^{-1} \ \rm{eV},
\label{eq:sync_peak}
\ea
\eeq
where $\gamma_{\min} \simeq \epsilon_e (m_p/m_e)(p-2)/(p-1)$
was used.
The Lorentz factor of electrons that cool on a time scale that is
equal to the dynamical time scale is $\gamma_c \sim 1$, thus above
$\varepsilon_{peak}$ the spectral index, $\nu F_\nu \propto
\nu^\alpha$, is $\alpha = 1 -p/2 = -1/2$, while below
$\varepsilon_{peak}$, $\alpha = 1/2$.
The self absorption frequency, $\varepsilon_{ssa}^{ob.} \simeq 100 \eV$, is
somewhat lower than the self absorption frequency predicted 
for a pure power law distribution of the electrons, 
\beq
\varepsilon_{ssa}^{ob.} =  600 \quad   
L_{52}^{2/3} \  \epsilon_{e,-0.5}^{1/3} \  \epsilon_{B,-0.5}^{1/3} \
\Gamma_{2.5}^{-8/3} \ \Delta t_{-2}^{-1} \ \rm{eV},
\eeq 
where a power law index $p=2$ for particles below $\gamma_{\min}$
is assumed.
This discrepancy is due to the fact that the low energy particles are
not power-law distributed, but have a quasi-Maxwellian distribution
due to photon reabsorption (see figure~\ref{fig:elec}).

Without pair production, the SSC model predictions of the Compton
scattering peak, at $\varepsilon_{IC,peak}^{ob.} = \gamma_{\min}^2
\varepsilon_{peak}^{ob.} = 1.3 \GeV$, agrees well with the numerical
result $\varepsilon_{IC,peak}^{ob.} = 1.5 \GeV$.  
The 1~GeV flux is comparable to the flux at 10~keV, as predicted by analytic
calculations based on the Compton $y$ parameter, $y=1$ in the scenario
presented in figure~\ref{fig:results1}. 

Pair production causes a cutoff at high energies.
For a flat spectrum, $\varepsilon^2 n_{ph}(\varepsilon)\propto
\varepsilon^0$ (which is a good approximation provided $\epsilon_B$ is
not much below equipartition), the optical depth to pair production is
well approximated by   
\beq
\ba{rcl}
\tau_{\gamma \gamma}(\varepsilon)& = & 
\Delta R n_{ph}(\varepsilon) {3 \over 16} \sigma_T  \nonumber \\ & = & 
\Gamma c \Delta t {U_{ph} \over \log 
\left( \frac{\varepsilon_{max}}{\varepsilon_{peak}}
\right)} {\varepsilon \over (m_ec^2)^2} {3 \over 16} \sigma_T,  
\ea
\eeq
and is larger than unity at 
\beq
\tilde\varepsilon^{ob.}_\tau \geq 3\times 10^8 
\log \left( \frac{\varepsilon_{max}}{\varepsilon_{peak}}
\right) \quad L_{52}^{-1} \epsilon_{e,-0.5}^{-1} \Gamma_{2.5}^6 \Delta t_{-2} 
\rm{\, eV},
\label{eq:epsilon_tau}
\eeq
in an excellent agreement with the numerical results.
Here $U_{ph}$ is the photon energy density,
given by $U_{ph} \approx \epsilon_e L / 4 \pi r_i^2 c \Gamma^2$. 
For this value of the compactness, pair annihilation does not play a
significant role, while scattering by the created pairs flattens the
spectrum at $10 \keV - 1 \GeV$.  

Even though the analytic approximation is in a fairly good agreement
with the numerical calculations, there are important discrepancies
between the analytic approximation and the numerical calculation.
The electron distribution shows a peak at $\gamma \sim 1.05$ ($\beta
\sim 0.3$), resulting in a deviation of the self absorption frequency
from the analytic calculation. These electrons affect the high energy
spectrum by Compton scattering, resulting in a nearly flat ($\alpha \sim 0$)
spectrum above 10~\keV. We showed \citep{Pe'er03b} that the spectrum
is nearly independent on the power law index of the accelerated
electrons, $p$.

\subsection{High compactness}
\label{sec:high_l}

\begin{figure}
\plotone{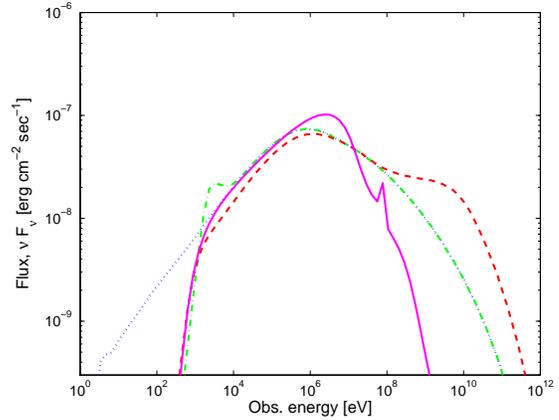}
\caption{Time averaged GRB prompt emission spectra obtained after 2
  shells collision, characterized by high compactness parameter (not
  corrected for the high optical depth to Thompson scattering). 
Results are shown for $L=10^{52}$~erg,
$\epsilon_e=\epsilon_B=10^{-0.5}$, $p=3$, $\Delta t = 10^{-4} {\rm \
s}$ , $\Gamma = 300$. The comoving compactness parameter is $l' = 250$.
Dotted curve: cyclo- synchrotron emission only. Dash dotted curve:
synchrotron emission and self absorption only. Dashed: synchrotron
emission, synchrotron self absorption and Compton scattering. Solid:
All processes included, including pair production and annihilation,
but excluding proton acceleration. 
Luminosity distance $d_L=2 \times 10^{28}$ and $z=1$ were assumed.
}
\label{fig:results2}
\end{figure}

Figure \ref{fig:results2} shows an example of our numerical results
for large comoving compactness, $l' = 250$.  At large value of the
compactness parameter, $l'>30$, the synchrotron-SSC model 
predictions do not provide an appropriate description of the spectrum.  
Therefore, the numerical results may provide some insight on the
 inconsistency between some of the observations and the analytical
  predictions, as mentioned in \S\ref{sec:intro}.

\begin{figure}
\plotone{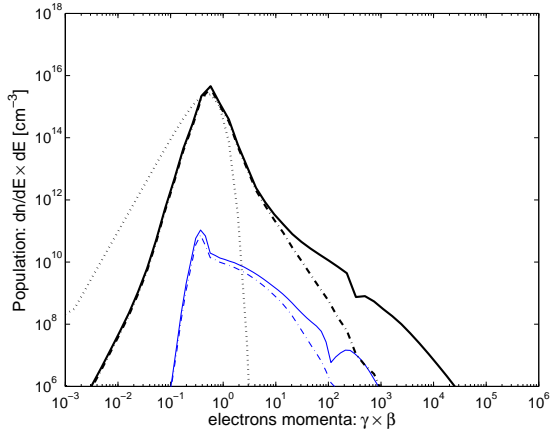}
\caption{Particle distribution at the end of the dynamical time.
Thick:  $\Delta t = 10^{-4} {\rm \ s}$, $l'=250$;
Thin: $\Delta t = 10^{-2} {\rm \ s}$, $l'=2.5$. 
All other parameters are the same as in Figure \ref{fig:results1}.
Solid: electron distribution, dash-dotted: positron distribution.
The dotted line shows a Maxwellian distribution at temperature 
$\theta\equiv kT/m_ec^2=0.08$.
}
\label{fig:elec}
\end{figure}

In the scenario of large compactness, Compton scattering by pairs
becomes the dominant emission mechanism. Both electrons and positrons
lose their energy much faster than the dynamical time, and a
quasi-Maxwellian distribution with an effective temperature $\theta
\equiv kT/m_ec^2 \simeq 0.05 - 0.1$ is formed. Photons upscattered by
the pairs create the peak at $\Gamma \theta m_e c^2 \sim 5 \MeV$.
The results shown in Figure \ref{fig:results2} are not corrected for
the fact that the optical depth to scattering by pairs is large,
$\tau_\pm \sim 10$ (see \S\ref{sec:procesess}). Therefore, the
emergent spectral peak is expect to be at lower energy, at $\sim
1\MeV$ \citep[for detailed discussion see][]{Pe'er03b}. 
The moderate Compton $y$ parameter, $y \simeq 4 \theta \tau \approx 4
\theta_{-1} \tau_1$ results in a spectral slope $\nu F_\nu \propto
\nu^\alpha$ with $\alpha \approx 0.5$ between $\varepsilon_{ssa} \approx
3 \keV$ and $\varepsilon_{peak} \approx 5 \MeV$. 
The peak at $\Gamma m_e c^2 \sim 10^2 \Gamma_{2.5} \MeV$ is formed
by pair annihilation. The self absorption frequency,
$\varepsilon_{ssa} \simeq 3 \keV$ is well below the prediction
for a power law index $p=2$ of particles below $\gamma_{\min}$. This
is attributed to the quasi-Maxwellian distribution of particles at low
energies (see Figure \ref{fig:elec}). The inverse Compton peak flux is
lower than the synchrotron peak flux, due to the Klein-Nishina
suppression at high energies. 

Even though the results presented here are for illustrative
 purposes only, and are not aimed to explain a particular
  observation, we note that the obtained numerical results are is
 agreement with some of the observations, that were found
 inconsistent with the optically thin synchrotron- SSC model
  predictions. Examples are the steep slopes observed at low energies
  \citep{Preece98, Frontera00, Ghirlanda03}, and the steep slopes
  above $\varepsilon_{peak}^{ob.}$ obtained by \citet{BB04}. 
Further results of our study are presented in
  \citet{Pe'er03b}. Comparison of the numerical results with the  
  high energy component reported by \citet{Gonzalez} are presented in
  \citet{PW1}.  

\section{Summary \& discussions}
\label{sec:summary}
\indent

We described a time dependent numerical model which calculates
emission of radiation from relativistic plasma, composed of homogeneous and
isotropic distributions of electrons, positrons, protons and photons,
and permeated by a time independent magnetic field. We assume the 
existence of a dissipation process, which produces energetic particles
at constant rates.
The particles interact via cyclo-synchrotron emission, synchrotron
self absorption, inverse and direct Compton scattering, $e^\pm$ pair
production and annihilation, and photo-meson interactions which
produce energetic photons and positrons following the decay of
energetic pions.
Exact cross sections valid at all energies, including the
Klein-Nishina suppression at high energies, are used in describing the
physical processes. Exact spectra are used  in describing
cyclo-synchrotron emission, synchrotron self absorption, Compton
scattering and pair production, and approximate spectra are used in
the description of pair annihilation.  

We explained in \S\ref{sec:numerical} our unique integration method,
which overcomes the challenge of the many orders of magnitude
difference in characteristic time scales.
We presented the various integration techniques used for solving the
kinetic equations describing the evolution of particle and photon
distributions at all energy scales. By following directly the 
development of the rapid, high energy electro-magnetic cascades at
each time step, we obtain the spectrum at high energies, up to $\geq
100 \TeV$. Our method enables to follow the development of the
spectrum created in the parameter space region of large compactness,
where no analytic approximation is valid.  
This method also improves over previous ones by providing a more
accurate treatment of photon emission and absorption in the presence
of magnetic fields.  

We have given several examples of numerical calculations in
\S\ref{sec:results}. In parameter space regions where analytical
approximations are valid, our numerical results are in good
agreement with analytic results. We have pointed out some significant
discrepancies between the analytical approximations and the numerical
calculation, and explained their origin. 
We presented examples of new results for parameter space regions where
analytic approximations are not valid.
We pointed out that our results are consistent with numerous
 observations, including observations that are inconsistent with the
  optically thin synchrotron-SSC model predictions.
 Further results of our study of
GRB prompt emission and early afterglow emission can be found in
\citet{Pe'er03b, Pe'er03}.  

The next generation high energy detectors, such as 
SWIFT\footnote{http://www.swift.psu.edu} and 
GLAST\footnote{http://www-glast.stanford.edu} satellites, and the 
sub-TeV ground based Cerenkov detectors, such as
MAGIC\footnote{http://hegra1.mppmu.mpg.de/MAGICWeb},   
HESS\footnote{http://www.mpi-hd.mpg.de/hfm/HESS/HESS.html}
VERITAS\footnote{http://veritas.sao.arizona.edu/} and 
CANGAROO~III\footnote{http://icrhp9.icrr.u-tokyo.ac.jp/} are
expected to increase the GRB prompt emission and early afterglow
emission detection rate by an order of magnitude, to allow detection
of $>\GeV$ emission from GRB's, and to detect the high energy spectra
of thousands of AGN's at various distances. 
Thus, detailed numerical models, that are capable of producing
accurate spectra over a wide energy scale, are necessary for analyzing
and understanding the experimental data.

\appendix
\section{Coulomb scattering}
\label{coulomb}
 In the limit of relativistic particle scattering off cool
thermal pair distribution ($\theta \ll 1$, $\gamma \gg1$),
the energy loss rate of the relativistic particle can be approximated
by 
$d\gamma/dt \approx -3/2 \sigma_T c n_{\pm} \ln \Lambda$, or
$t_{ee}^{-1} \equiv - (1/\gamma) d \gamma /dt \approx  
4 \pi c r_0^2 n_{\pm} \gamma^{-1} \ln \Lambda$ \citep{Gould75},
where $n_{\pm}$ is the number density of the thermal pairs.
The relevant value of the Coulomb logarithm $\Lambda$, is 
$\Lambda \approx \gamma^{1/2} m_e c^2 / h \omega_p$, where 
$\omega_p = (4 \pi n_{\pm} e^2 / m_e)^{1/2}$ is the plasma
frequency.
 
Assuming that the pairs' energy distribution is thermal, that their
number density is $n_{\pm} \equiv f n_p$ where $n_p$ is the proton
number density, and that $n_p \approx u_{int}/m_p c^2$ where $u_{int}$
is the internal energy density, comparing the Coulomb cooling time
and the synchrotron cooling time, $t_{syn} = 9 m_e^3 c^5 / 4 q^4 B^2
\gamma$, using  $B^2 = 8 \pi \epsilon_B u_{int}$, gives 
\beq
{t_{syn} \over t_{ee}} = {9 \over 8} {m_e \over m_p} 
\frac{f \ln \Lambda}{ \epsilon_B \gamma^2} \simeq
\frac{1}{3\gamma^2} \, f_1 \epsilon_{B,-0.5}^{-1},
\eeq
where typical values $f = 10 f_1$, and $\ln \Lambda \approx 20$
assumed.
It is therefore concluded that for relativistic electrons, and for
magnetic field not many orders of magnitude below equipartition,
electrons lose their energy by synchrotron emission on a time scale
much shorter than the energy loss time by Coulomb scattering. A more
accurate approximation of $t_{ee}$ \citep{Haug88, Coppi90}, does not
change this result.

\end{document}